# QUANTUM THREE-PASS PROTOCOL: KEY DISTRIBUTION USING QUANTUM SUPERPOSITION STATES

Yoshito Kanamori[1] and Seong-Moo Yoo[2]

[1] Department of Computer Information Systems, University of Alaska Anchorage, Alaska, USA

yoshitok@cbpp.uaa.alaska.edu

[2] Electrical and Computer Engineering Department, The University of Alabama in Huntsville, Alabama, USA

yoos@eng.uah.edu

*ABSTRACT*

*This letter proposes a novel key distribution protocol with no key exchange in advance, which is secure as the BB84 quantum key distribution protocol. Our protocol utilizes a photon in superposition state for single-bit data transmission instead of a classical electrical/optical signal. The security of this protocol relies on the fact, that the arbitrary quantum state cannot be cloned, known as the no-cloning theorem. This protocol can be implemented with current technologies.*

## 1. INTRODUCTION

Research on quantum computation has been getting more and more attention for the last two decades. It is known that, once a quantum computer is built, the existing popular public-key encryption algorithms (e.g., RSA, elliptic curve cryptography) may be compromised in polynomial time [1] [2]. Therefore, cryptographic schemes which do not rely on computational complexity, have been expected. Quantum cryptography is one, that is theoretically unbreakable and does not rely on computational complexity, but on quantum mechanical properties.

Although many quantum cryptographic schemes have been proposed [3] [4] [5] [6], the one well researched and realized experimentally [7] [8] [9] [10] [11] is the quantum key distribution protocol (QKD). Also, some QKD commercial products are available [12] [13]. The QKD schemes, in general, utilized photons to transfer classical bit information. For example, in the BB84 protocol proposed by Bennett and Brassard in 1984 [14], a sender (Alice) chooses one of two different orthogonal base sets (i.e., {horizontal, vertical} or $\{+45, -45\}$) and sends one of two polarization states in the chosen base. The receiver (Bob) measures the received photon with the one chosen randomly out of the two base sets. After a certain amount of photons have been transmitted, Alice and Bob exchange the information, by using a classical channel, about the base sets that they used for polarizations and measurements so that they can share the same data bits. In 1991, Ekert proposed a key distribution protocol using the Einstein-Podolsky-Rosen (EPR) pair and Bell's inequality [15]. In1992, Bennett simplified BB84 using only one non-orthogonal base [16]. Since both a sender and a receiver will randomly choose one of the two orthogonal base sets, the processes of generating a key bit sequence in these three QKD protocols are nondeterministic. Also, lots of transmitted data will be discarded during the process of sifting through data bits in these QKD systems. Recently, deterministic key distribution protocols are proposed. Quantum secure direct communication (QSDC) is implemented by exchanging single photons with classical channel [17]. Ping-pong QSDC uses EPR pairs [18]. The weaknesses of the ping-pong scheme were found by Wójcik [19] and Cai [20]. They improved the ping-pong scheme. QSDC with a one-time pad is also proposed [21] [22]. Quantum dense key distribution utilizes QKD and quantum dense coding [23] to implement the key distribution enhancing the capacity of transmission [24] [25] [26].

In this paper, we propose a quantum key distribution protocol utilizing a quantum superposition state, called Quantum Three Pass Protocol (QTTP) [27]. This protocol provides the same security as BB84 does. The security relies on the fact that the arbitrary quantum state cannot be cloned [28], known as the no-cloning theorem [29]. The QTPP requires neither key-exchange in advance, nor communication on classical channels, unlike the QSDC. There are three main advantages compared to BB84. First, since the QTPP is a deterministic secure communication protocol, all data bits received from a sender can be used as encryption keys ideally, while BB84 needs to discard half of the transmitted data because the probability that both Alice and Bob use the same orthogonal bases is 50 % [30]. Since the data bit sequence to be transmitted is determined in advance, existing error correcting codes can be applied to the data bits in order to reduce transmission errors. Second, the QTPP system can detect the existence of eavesdroppers even under severely noisy environments where BB84 cannot detect them. Third, as long as the measurement base stays still during the session, the QTPP does not require as precise base alignments as does BB84 to encrypt quantum states.

This paper is organized as follows: In section 2, we introduce a classical three-pass protocol. In section 3, we present our protocol QTPP, and the security analysis is presented in section 4. Finally, conclusions are presented in section 5.

## 2. THREE-PASS PROTOCOL

Let us consider a figurative example of our three-pass protocol. There is a way to deliver a message in a box securely to a receiver using two padlocks, without sharing keys to open the locks. First, Alice locks the box with her padlock and sends the box to Bob. Bob receives the box and locks the box with his padlock and sends the box to Alice. At this point, the box has two padlocks. Then, Alice takes off her padlock from the box and sends it again to Bob. Bob removes his padlock from the box. Bob got the message securely from Alice. If we use this idea in cryptography, the algorithm establishes secure communication with no key-exchange in advance.

This protocol may be realized by utilizing exclusive-OR (XOR) operations. First, Alice performs XOR of a message $M$ with her key $K_A$ and sends the result $M \oplus K_A$ to Bob. ('$\oplus$' denotes XOR operation.) Bob performs XOR of $M \oplus K_A$ with his key $K_B$ and sends $M \oplus K_A \oplus K_B$ to Alice. Alice performs XOR of the received massage $M \oplus K_A \oplus K_B$ with her key $K_A$ and sends the resulting message $M \oplus K_B$ to Bob. Bob performs XOR of $M \oplus K_B$ with his key $K_B$ and gets $M$ from Alice securely. This interesting protocol seems to allow us to communicate each other securely without sharing the secret keys. However, there is a fatal weakness in this scheme. If Eve (who eavesdrops on the communication) can make a copy of all three messages exchanged between Alice and Bob, Eve can retrieve the original message $M$ by simply performing XOR of all three messages as follows:

$$(M \oplus K_A) \oplus (M \oplus K_A \oplus K_B) \oplus (M \oplus K_B) = M$$

Therefore, this three-pass protocol with simple XORs does not work. Shamir's three-pass protocol (never published but introduced by Massey [31]) uses a discrete logarithm problem instead of a simple XOR operation to avoid the above weakness, but it has never been proved [32]. Shamir's protocol requires encryption keys that are correlated to each other. Thus, it is no longer a protocol with no key-exchange in advance. The problem of the three-pass protocol with XORs comes from the fact, that an eavesdropper can make copies of the transmitted messages. As long as a classical computer uses a classical signal (e.g., electrical or optical signal) for transmissions, the three-pass protocol seems to be infeasible since eavesdroppers can easily make copies of the transmitted data and analyze them.

However, Quantum Mechanics makes it feasible to implement this three-pass protocol without key-exchange in advance. The QTPP utilizes a particle (e.g., photon) in quantum superposition states for one bit transmission, instead of a classical signal.

## 3. QUANTUM THREE-PASS PROTOCOL (QTPP)

Unlike the BB84 protocol, QTPP requires only quantum channels. Classical one-bit information is encoded into a single particle, called a "quantum bit" or "qubit" whose state is represented by using a vector (e.g., $|0\rangle$ or $|1\rangle$,) called "ket" vector in Dirac notation. In this paper, we assume that a photon is used as a qubit. (Henceforth, we use photon and qubit interchangeably.) We use a photon as a qubit and one polarization base set {horizontal, vertical} to represent a classical two level system. A horizontally polarized photon represents logic zero, $|0\rangle = (1 \; 0)^T$ and a vertically polarized photon represents logic one, $|1\rangle = (0 \; 1)^T$. When a sender's message $M$ has $n$ classical bits, the encoded qubit states can be represented as $M = |i_1\rangle \otimes |i_2\rangle \otimes |i_3\rangle \otimes ... \otimes |i_j\rangle$, where $\{i_j \, | \, i_j = 0 \; or \; 1, j = 1, 2, ..., n\}$. '$\otimes$' represents a tensor product. After an $n$-bit message is encoded into $n$ photons, the polarization of each photon is rotated by an angle $\theta_j$, which is chosen randomly for each qubit. The rotating operation can be considered as an encryption and represented as

$$R(\theta_j) = \begin{pmatrix} \cos\theta_j & \sin\theta_j \\ -\sin\theta_j & \cos\theta_j \end{pmatrix}.$$

The angle $\theta_j$ can be considered as an encryption key. (Decryption can be performed by rotation with the angle $-\theta_j$.) Since each qubit requires a different angle, this encryption is similar to a classical one-time pad which is a perfect encryption scheme in classical cryptography [32]. However, in a QTPP system, the key is not shared between users unlike the one-time pad system. Alice and Bob generate their own secret key $K_A$ and $K_B$ ($K = \{\theta_j \, | \, 0 \leq \theta_j < \pi, j = 1, 2, ..., n\}$) for each session. Those session keys are never disclosed to anyone. Each session key is used only twice in the session by the generator: once for encryption and once for decryption. After each session, the key is discarded and a new key is generated to prevent any information related to the session key and data from being leaked.

In the following discussion, without losing generality, we can assume that message $M$ is single photon encoded as $M = |0\rangle$ (i.e., $n = 1$ and $i_1 = 0$) and Alice initiates a key distribution (Refer to Figure 1). First, Alice and Bob generate their session keys $K_A = \theta_A$ and $K_B = \theta_B$. Alice encrypts $M$ with her encryption key $K_A$. The resulting state can be described as

$$E_{K_A}[M]: \quad R(\theta_A)|0\rangle = \begin{pmatrix} \cos\theta_A & \sin\theta_A \\ -\sin\theta_A & \cos\theta_A \end{pmatrix}\begin{pmatrix} 1 \\ 0 \end{pmatrix} = \cos\theta_A \cdot |0\rangle - \sin\theta_A \cdot |1\rangle = |\psi_1\rangle,$$

where $E_{K_A}$ indicates an encryption with $K_A$. Such a resulting state is called a superposition state. Alice sends the resulting state $|\psi_1\rangle$ to Bob. Bob receives the photon in $|\psi_1\rangle$ and encrypts it with his key $K_B$.

$$E_{K_B}[E_{K_A}[M]]: \quad R(\theta_B) \cdot |\psi_1\rangle = \cos(\theta_B + \theta_A)|0\rangle - \sin(\theta_B + \theta_A)|1\rangle = |\psi_2\rangle.$$

The resulting state $|\psi_2\rangle$ is still a superposition state. Bob sends it back to Alice. Alice receives and decrypts it by rotating it back with the angle $\theta_A$ (i.e., rotation of $-\theta_A$) and sends the resulting superposition state $|\psi_3\rangle$ to Bob.

$$D_{K_A}[E_{K_B}[E_{K_A}[M]]] = E_{K_B}[M]: \quad R(-\theta_A) \cdot |\psi_2\rangle = \cos\theta_B \cdot |0\rangle - \sin\theta_B \cdot |1\rangle = |\psi_3\rangle,$$

where $D_{K_A}$ indicates an decryption with $K_A$. Bob receives and decrypts it by rotating it back with the angle $\theta_B$ (i.e., rotation of $-\theta_B$).

$$D_{K_B}[E_{K_B}[M]]: \quad R(-\theta_B) \cdot |\psi_3\rangle = \begin{pmatrix} \cos(-\theta_B) & \sin(-\theta_B) \\ -\sin(-\theta_B) & \cos(-\theta_B) \end{pmatrix} \cdot \begin{pmatrix} \cos\theta_B \\ -\sin\theta_B \end{pmatrix} = \begin{pmatrix} 1 \\ 0 \end{pmatrix} = |0\rangle.$$

Now, Bob has the original message $M = |0\rangle$.

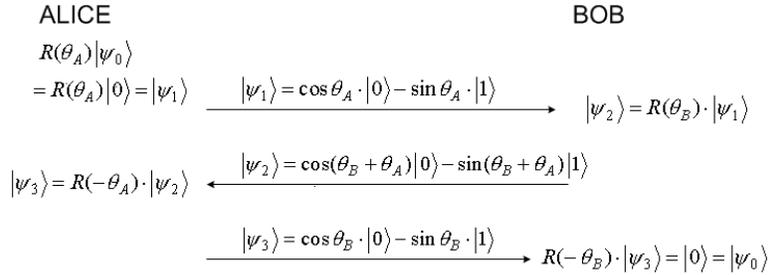

Figure 1: QTPP protocol example

## 4. SECURITY OF QTPP

As mentioned earlier, the classical three-pass protocol with XORs does not work, because Eve (an eavesdropper) can copy the transmitted data and retrieve the plain message. In order to prevent Eve from copying the data bits, the QTPP uses a particle in quantum superposition state as the media of the one-bit transmission. In a quantum system, the replication of arbitrary quantum state (i.e., $a|0\rangle + b|1\rangle$, $a$ and $b$ are arbitrary coefficients) is infeasible [28] [29]. In short, Eve cannot copy the transmitted data encoded in superposition states without errors. Eve may try to apply intercept-resend attack [33] instead of copying the data. However, the QTPP is also secure against the intercept-resend strategy. When Eve intercepts a photon that was in logic zero state (i.e., $|\psi\rangle = |0\rangle$) before it was encrypted, the state of the received photon can be represented as $|\psi\rangle = \cos\theta|0\rangle - \sin\theta|1\rangle$. The angle $\theta$ is always unknown to Eve. Both "$\cos\theta$" and "$-\sin\theta$" are the probability amplitude of the quantum state. If Eve tries to read the data in this photon, she will observe $|0\rangle$ with probability $|\cos\theta|^2$ and $|1\rangle$ with probability $|-\sin\theta|^2$. When Eve intercepts a photon that was in logic one state (i.e., $|\psi\rangle = |1\rangle$) before it was encrypted, the state of the received photon can be represented as $|\psi\rangle = \sin\theta|0\rangle + \cos\theta|1\rangle$. Therefore, as far as the classical data bits (i.e., logic zero and one) are encoded randomly into photons, Eve's measurement result will be $|0\rangle$ with probability $\frac{1}{2}|\cos\theta|^2 + \frac{1}{2}|\sin\theta|^2$ and $|1\rangle$ with probability $\frac{1}{2}|-\sin\theta|^2 + \frac{1}{2}|\cos\theta|^2$. When we use the QTPP for key distribution, the message data is a sequence of random bits. As a result, Eve will observe $|0\rangle$ or $|1\rangle$ with probability $\frac{1}{2}$. Therefore, the QTPP guarantees the confidentiality of communication. Eve cannot resend the intercepted data to Bob without errors.

Also, the existence of Eve can be detected, because Eve's intercept-resend attack increases the bit error rate up to 50%. This bit error rate, caused by eavesdropping, is much higher than the one of the BB84 (i.e., 25% [30]). In short, eavesdroppers can be detected more easily in the QTPP system than a system with the BB84 protocol. Moreover, when the error rate is significantly increased, higher than 25 % because of severe noisy environments, the BB84 protocol system no longer distinguishes between the transmission error caused by Eve and the one caused by innocent noisy environment, while the QTPP system is still able to detect the existence of Eve under such a circumstance, unless the error rate becomes close to 50 %.

Another possible attempt to get information from the transmitted data is that, Eve may entangle the intercepted data with Eve's qubit, which is called individual particle attack [33]. However, Eve will still observe $|0\rangle$ or $|1\rangle$ randomly. For example, Eve can apply controlled-NOT [29] operation to the transmitted data as the control bit and her qubit $|0\rangle$ and $|1\rangle$ as the target bit. In this case, Alice and Bob cannot detect the eavesdropping since this operation does not change the state of the transmitted qubit. Using this technique, Eve can entangle her three qubits with three transmitted data qubits (i.e., Alice→Bob, Bob→Alice, Alice→Bob) and apply unitary operations with her three qubits in order to try to get the plain message. However, Eve cannot retrieve the plain data bit, because Eve does not know the states of entangled particles and the angles of Alice and Bob.

## 5. CONCLUSION

This paper proposed a quantum three-pass protocol (QTPP). This protocol makes use of the advantage of the classical three-pass protocol and the advantage of quantum mechanical property. Since this protocol uses all transmitted data unlike the BB84, it can be used not only for deterministic QKD, but also for data transmission utilizing classical existing error correcting codes. The security against the known and practical feasible attacks was discussed. The QTPP can detect eavesdroppers under a severe noisy environment where the BB84 may not be able to detect eavesdroppers. Also, as long as the measurement base stays still during the session, the QTPP does not require as precise base alignments as the BB84 does because the slight discrepancy of polarization angles caused by misalignments can be included into part of a key angle.

Just like the BB84 protocol [34], the QTPP requires an authentication before the key distribution process starts. A simple authentication can be implemented by sharing a secret bit sequence between participants as a key and performing XOR of the shared key with a message (e.g., key to be distributed) prior to the key distribution process introduced in this paper. Moreover, the QTPP scheme can be modified for other cryptographic schemes (e.g., authentication, secret sharing).

Even though polarization encoding is used in this paper to make a superposition state, phase encoding would be also applied to the QTPP. This protocol can be implemented with the current technologies (i.e., single photon source and detector, mirror and faraday rotator) and is a good example of how an infeasible protocol in a classical computer system can be realized with quantum mechanical resources.